\documentclass[a4paper]{jpconf}

\usepackage{graphicx}

\usepackage{amssymb}
\usepackage{mathtools}
\usepackage{tikz}

\newcommand\ee{\end{equation}} 
\newcommand\be{\begin{equation}}

\newcommand\mathC{\mkern1mu\raise2.2pt\hbox{$\scriptscriptstyle|$}
        {\mkern-7mu\rm C}}              

\def\be{\begin{equation}}
\def\ee{\end{equation}}
\def\bear{\begin{eqnarray}}
\def\eear{\end{eqnarray}}

\def\ket{\rangle}

\begin{document}
\title{Non-locality and quasiclassical reality in Kent's formulation of relativistic quantum theory}

\author{J Butterfield$^1$, B Marsh$^2$}

\address{$^1$ Trinity College, Cambridge CB2~1TQ, UK}
\address{$^2$ SITP and Department of Applied Physics, Stanford University, Stanford CA 94305, USA}

\ead{jb56@cam.ac.uk, marshbp@stanford.edu}

\begin{abstract}
We report Adrian Kent's proposed framework for a realist, one-world, Lorentz-invariant formulation of quantum theory. The idea is to postulate a final boundary condition: in effect, a late-time
distribution of mass-energy recording how photons scattered off macroscopic objects. Nature selects this final boundary condition with the orthodox late-time Born probability; and this defines the probability space of events, to give a realist quantum theory. We emphasize two topics. First, we consider this formulation's verdicts about traditional locality conditions, such as Outcome Independence and Parameter Independence. Second, we discuss a possible amendment to Kent's proposal that, roughly speaking, allows for the emergence of a quasiclassical history even when mass-energy is shielded or delayed from appearing in the final boundary condition.
\end{abstract}

\section{Introduction}
Kent proposes a solution to quantum theory's measurement problem \cite{MainKent,ToyModels,PWM} that is realist, one-world, and relativistic, in contrast to `the usual suspects' of e.g. Bell's `Six possible worlds of quantum mechanics' \cite{Bell}. The idea is to postulate a final boundary condition: in effect, a late-time distribution of mass-energy recording how photons scattered off macroscopic objects. Nature selects one such final boundary condition with the late-time Born probability, and so the probability space of events is defined. Given the final boundary condition, a Lorentz-invariant rule defines a generalized field of beables in spacetime describing the quantum reality of the system. We will: (i) set the proposal in the context of quantum non-locality; (ii) explore a cousin proposal with a different Lorentz-invariant rule used to formulate beables. Our discussion will summarize and develop our recent work \cite{Butterfield,Marsh2018,Marsh2019} in this area.

\section{Peaceful coexistence?}\label{PC?}
Thirty years ago, Shimony hoped to reconcile quantum non-locality with special relativity---to secure `peaceful coexistence'---by carefully isolating the culprit in the derivations of the Bell inequality. The main assumption of a Bell inequality is factorizability:
\be
\forall \lambda; \forall x,y; \forall X, Y = \pm 1: \;\;
\textrm{Pr}_{\lambda, x,y}(X \& Y) =  \textrm{Pr}_{\lambda, x}(X) \cdot \textrm{Pr}_{\lambda,  y}(Y) \; ;
\label{facby}
\ee
where $\lambda$ encodes a complete state, $x$ and $y$ are choices of possible measurements in the left and right wing of a Bell experiment, and $X$ and $Y$ possible outcomes. Factorizability is the conjunction of two conditions, Parameter Independence (PI) and Outcome Independence (OI):	
\be
\forall \lambda; x,y; X, Y = \pm 1: \;\;
\textrm{Pr}_{\lambda, x}(X) = \textrm{Pr}_{\lambda, x,y}(X) :=   \textrm{Pr}_{\lambda, x,y}(X \& Y) + \textrm{Pr}_{\lambda, x,y}(X \& \neg Y) \; ;
\label{pi}
\ee
\be
\forall \lambda, x,y; X, Y = \pm 1: \;\;
\textrm{Pr}_{\lambda, x,y}(X \& Y) =  \textrm{Pr}_{\lambda, x,y}(X) \cdot \textrm{Pr}_{\lambda, x,y}(Y) \; .
\label{oi}
\ee

With $\lambda$ as the quantum state, PI ensures that the statistics in one wing are not correlated with which quantity is chosen to be measured in the other wing. So for spacelike measurement events, PI ensures the impossibility of superluminal signaling (the information in such a signal being which quantity has been chosen to be measured). This is essentially the content of the no signalling theorem, which is derived in quantum theory from the commutation of quantities in the left wing with those on the right. So Shimony suggested that to get peaceful coexistence, it is enough to deny OI \cite{Shimony1986,Shimony2009}. 

On the other hand, consider the verdicts about PI and OI made by the pilot-wave theory. In its usual version, we take $\lambda$ as the conjunction of the quantum state and the particles' possessed positions. OI holds thanks to the theory's determinism. But PI fails since the velocity of each particle  depends instantaneously on the position of the other, thanks to ${\bf p}_i = \nabla_i S$, where $S \equiv S({\bf x}_1, {\bf x}_2)$ is the phase of the wave function in configuration space. Since the pilot-wave theory (in this version) is usually considered incompatible with special relativity, these two verdicts seem to support Shimony's contention that denying OI and endorsing PI secures peaceful coexistence.

But we submit that one should not make a fundamental distinction between parameters (i.e. apparatus-settings) and outcomes (i.e. measurement results).  For they are both macrophysical facts: so presumably, they are on equal terms, as regards whether a curious (i.e. unscreenable-off) correlation between examples of them at spacelike separation violates relativity theory. This seems to be the viewpoint of John Bell, especially in his last essays, `Against measurement' \cite{AgainstMeasurement} and `La nouvelle cuisine' \cite{NouvelleCuisine}.  As Bell might have put it: `surely Nature, in her causal structure, does not care whether a macrophysical fact is `controllable' by humans, in the way a knob-setting, but not a pointer-reading, is---or at least seems to be?' More generally, the fact that the parameter/outcome distinction in PI and OI is schematic reinforces the point that having verdicts about PI and OI sufficient to avoid a Bell inequality does not amount to having a solution to the quantum measurement problem. So proposed solutions to that problem still need to be addressed. 


For assessing Kent's proposal (in Section 3 et seq.), it will be important to recall that Bell's theorem predicts observable probabilities by averaging over $\lambda$. For example, letting $a_1$ be a left-wing setting (i.e. a value of $x$) and $A_1$ the corresponding outcome:
\be\label{recoverbyaverage}
\textrm{Pr}(A_1 = + 1)  := \int_{\Lambda} \; \textrm{Pr}_{\lambda, a_1}(+ 1) \; d \rho \; .
\ee
Here, the fact that $\rho$ is independent of which quantity, e.g. $a_1$ vs. $a_2$, is  measured, encodes a locality assumption. Namely, there is no correlation between

\begin{enumerate}
    \item the causal factors influencing which quantity, e.g. $a_1$ vs. $a_2$, is measured, and
    \item the causal factors influencing which value of $\lambda$ is realized.
\end{enumerate}

This {\em no conspiracy} assumption is very reasonable  if: (i')  the causal factors (i) are localized in the wings of the experiment, and (ii') the causal factors (ii) are localized in the central source. But with an eye on our later discussion of Kent's proposal, we should ask: what if $\lambda$ encodes facts about a final boundary condition? In particular: what if $\lambda$ encodes information in that boundary condition about traces (records) of earlier choices of measurement? The answer is surely that, if so, a probability distribution over values of $\lambda$ may well depend on such choices---so that the label `no conspiracy' is tendentious, since such dependence is in no way suspicious. 

\section{Kent's formulation of quantum theory}\label{AK}

Kent's proposal is realist, one-world, and without state reduction. In those three features, it is like the pilot-wave theory. But it is relativistic, and its beables are very different from those of the pilot-wave theory. 

Kent's main idea can be stated for Minkowski spacetime, as follows. He takes the universal quantum state to evolve unitarily, and to determine Born-rule probabilities for the outcomes of a {\em hypothetical} mass-energy density measurement at all points of an asymptotic late-time hypersurface $S$. Then he supplements the orthodox description of the quantum system given by the state, with one such mass-energy density distribution---one which ``Nature selects''  at random, with Born-rule probabilities prescribed by the universal quantum state at the late ``time'' $S$. Then he proposes a Lorentz-invariant rule that invokes orthodox quantum correlations with whichever distribution of mass-energy density on $S$ ``Nature has selected'', so as to define beables at points of spacetime earlier than $S$. Thus Kent supplements quantum theory with asymptotic mass-energy distributions so as to define beables---facts about quantities' values going beyond quantum orthodoxy---at earlier points of spacetime. 

The physical idea of his rule defining the beables is decoherence. In short: that a photon registers on $S$ in one place rather than another---as encoded in the randomly selected mass-energy distribution on $S$---is correlated with its having earlier scattered off some macroscopic object in one way rather than another: and this scattering typically encodes the location of the macroscopic object. Thus the selected mass-energy distribution on $S$ acts as a record of facts, like macroscopic objects' locations, at earlier times. The idea is that enough photons (or other particles) register, so as to record many definite locations of objects at earlier times: sufficiently many to imply an intuitively satisfying macroscopic world, or `quasiclassical realm'---and so to give a realistic and one-world solution of the measurement problem.  Besides, the fact that the photons travel along null (lightlike) curves makes the correlations encoded in these records suitably Lorentz-invariant.

This overview of Kent's proposal is summarized in Figure 1.

\begin{figure}[h]
    \center
    \includegraphics[scale=0.8]{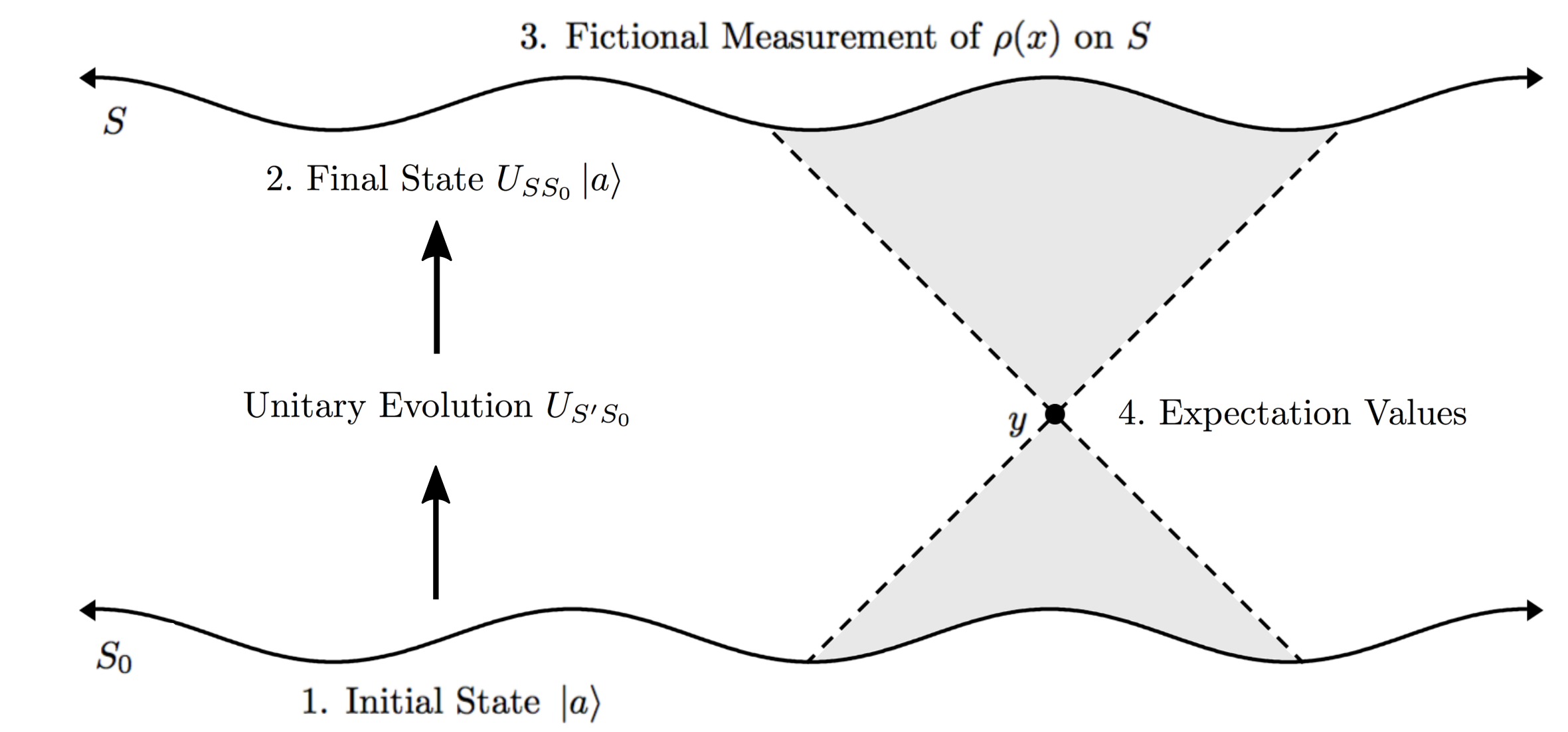}
    \caption{Overview of Kent's proposal for deriving beables for any spacetime point $y$ between an initial spacelike hypersurface $S_0$ and a final spacelike hypersurface $S$. 1. An initial quantum state is provided on an initial spacelike hypersurface $S_0$. 2. The state unitarily evolves to any spacelike hypersurface $S'$ in the future of $S_0$ via an appropriate unitary operator $U_{S'S_0}$ provided by the Tomonaga-Schwinger formalism \cite{TomonagaSchwinger}. 3. The final state $U_{SS_0}|a\ket$ on $S$ determines the Born rule probability for any given final boundary condition. 4. A single final boundary condition is used in a Lorentz-invariant rule to compute beables at spacetime point $y$, using only the part of the final boundary condition {\em outside} the future light cone of $y$. }
    \label{OverviewFig}
\end{figure}

We should add that this overview has simplified in two related respects. First: Kent's proposal does not require Minkowski spacetime. He discusses how it can be transcribed to a large range of cosmologies, that admit late hypersurfaces on which---one can reasonably suppose---enough photons (or other particles) register, so as to record sufficiently many definite locations of objects that one intuitively wants (in a realistic, one-world solution of the measurement problem) to be definitely located. But in this paper, we can keep Minkowski spacetime in mind, as our main example. Second: the proposal does not need to postulate some single hypersurface $S$, the various possible mass-energy distributions on which are then the main ingredient for defining the extra beables: i.e. the facts that supplement quantum orthodoxy, analogous to the corpuscles' positions according to pilot-wave theory. After all, what matters for a realistic one-world solution of the measurement problem is a well-defined probability space of  events: just {\em one} of which happens. Thus Kent envisages defining such a space by a sequence of ever-later hypersurfaces and their associated probability distributions: i.e. by a convergent sequence of probability distributions. But for simplicity of exposition, we will always talk about just one such late hypersurface $S$.

To sum up: Kent proposes to solve the measurement problem by a notional detection of photons at a late time. We are to think of a notional L\"{u}ders-rule measurement of mass-energy density at each point on $S$. The measured outcome distribution: (i) is correlated with where and when photons scattered off macroscopic objects at earlier times, and so: (ii) defines the probability space of real events. 

In Section 3.1 et seq., we will illustrate these ideas with two of Kent's toy models of photons propagating to $S$ after scattering off an object. But first we further clarify the proposal in general terms, with three numbered comments.

\subsection*{1. Supplementing decoherence}

 Recall the main idea of decoherence. It is not so much the destruction of coherence (i.e. of quantum probability distributions' interference terms that are characteristic of a superposition as against a mixture), but rather the {\em diffusion} of coherence from the system to its environment. 
 
 Thus in the paradigm case studied in e.g. \cite{Schlosshauer} of photons scattering off the pointer of a macroscopic apparatus,  the reduced state (density matrix) of the pointer very rapidly becomes nearly diagonal in a quantity, such as position of its centre of mass, which---to secure a definite `quasiclassical' macroscopic world---one intuitively wants to be definite in value. But decoherence does not select a single component of the reduced state / density matrix of the system. Nor does it justify attributing a vector state to the system. For the reduced state cannot be interpreted as a matter of ignorance. 
 
 This last point is independent of the physical process of decoherence; and is of course, the source of the measurement problem, i.e. quantum orthodoxy's lack of  values for quantities, e.g. macroscopic positions, that one intuitively wants to be definite. It was emphasized long ago by D'Espagnat, who suggested calling such reduced states {\em improper mixtures}, as against the straightforwardly ignorance-interpretable proper mixtures describing e.g. an impure beam of electrons prepared by a less-than-perfect electron-gun \cite{Espagnat}. 
 
 Indeed, it was emphasized {\em longer} ago by Schr\"{o}dinger in his great 1935 papers, that introduced his famous cat and the word `entanglement'. Thus in the first paper, `The present situation in quantum mechanics' \cite{schrodinger1935,trimmer1980}, he says this `is the most difficult and most interesting point of the theory'. He summarizes as follows: `Maximal knowledge of a total system does not necessarily include total knowledge of all its parts, not even when these are fully separated from each other and at the moment
are not influencing each other at all. ...  Best possible knowledge of a whole does not necessarily include the same
for its parts. ... The whole is in a definite state, the parts
taken individually are not.' And he reinforces the point in a dialogue with an imagined interlocutor:\\
\indent  ``How so? Surely a system must be in some sort
of state." \\
\indent ``No. State is $\psi$-function, is maximal sum
of knowledge. I didn't necessarily provide myself
with this, I may have been lazy. Then the system is
in no state."\\
\indent ``Fine, but then too the agnostic prohibition of
questions is not yet in force and in our case I can
tell myself: the subsystem is already in some state, I
just don't know which."\\
\indent ``Wait. Unfortunately no. There is no `I just don't know'. For as to the total system, maximal
knowledge is at hand . . " (Section 10, pages 331-332, of \cite{trimmer1980}).


\subsection*{2. The light cone is respected} 

In Kent's proposal, the physical beables at spacetime points $y$ in the past of $S$ are determined only by the part of the final boundary distribution on $S$ that is outside the future light cone of $y$: `the light cone is respected'. Restriction to only the spacelike-separated region of the final boundary condition yields a Lorentz-invariant rule for the construction of the beables. Cf. Figure 2. (Here, `determined' is to be understood in a tenseless sense: there is no backwards causation from $S$ to $y$.)

\begin{figure}[h]
\includegraphics[scale=0.8]{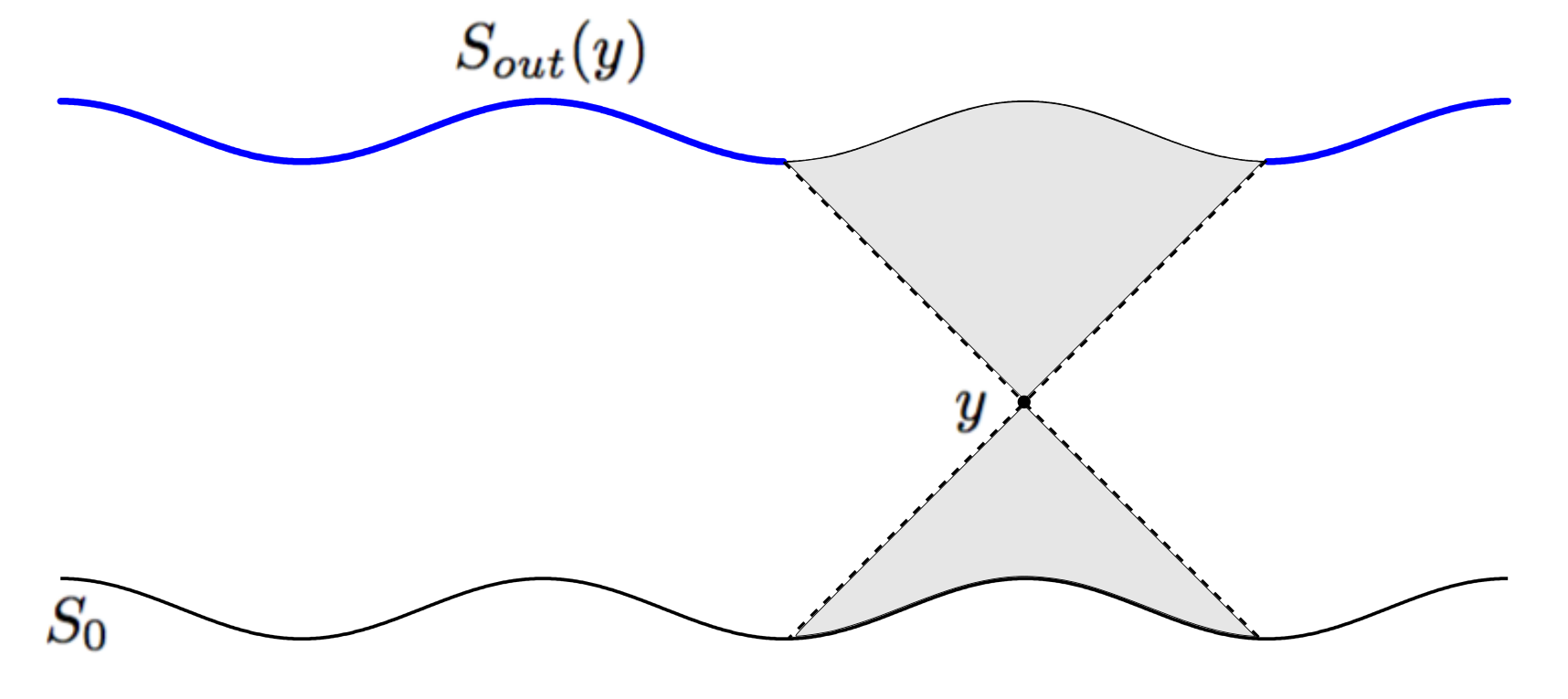}\hspace{1pc}%
\begin{minipage}[b]{18pc}\caption{The spacelike hypersurface $S_{out}(y)$ in bold blue, containing the part of the final boundary condition used to determine beables at $y$.}
\end{minipage}
\end{figure}

\subsection*{3. Conditioning on a final boundary condition}

We emphasize that although conditioning on a final boundary condition is of course much less common in physics than conditioning on an initial condition, it is in no way  suspicious. In particular, it is a matter of logical relations between physical statements, not of causation: so it does not imply any backwards causation.

Indeed, the legitimacy of considering final as well as initial conditions is illustrated by various time-symmetric formalisms for quantum theory: in particular, by the well-known ABL rule \cite{aharonov1964}, which gives the probability, conditioned on both final and initial states of a non-relativistic quantum system, of results of L\"{u}ders-rule  measurements at intermediate times.

In a bit more detail: a non-relativistic system is prepared in an eigenstate $|{\psi(0)}\ket = | {a} \ket$ of observable A. At a later time $T$, observable C is measured and the eigenvalue $c$ with eigenstate $|{c}\ket$ is found. We consider an intermediate measurement of observable B with eigenvalues $\{b_i\}$, occurring at time $0< t<T$. That is, we ask: what is the probability of obtaining outcome $b_i$ at intermediate time $t$, conditioned on (i) the initial state $| {\psi(0)} \ket =|{a}\ket$ \textit{and} on (ii) obtaining outcome $c$ at the final time $T$?

The conditional probabilities for intermediate measurement outcomes follows as a direct result of the measurement postulate and the Born rule. For a given initial state $| {a} \ket$ and final measurement outcome $c$, the conditional probability $\textrm{Pr}(\,b_i\,\bigm\vert\, c , |{a}\ket)$ of measuring $b_i$ is prescribed by the ABL rule---see \cite{Marsh2018} for a detailed derivation. Precisely this ABL probability is used to construct the beables of Kent's non-relativistic proposals, as in Section two of his paper \cite{MainKent}.

Agreed: Kent's relativistic proposal does not use an exact relativistic generalization of the ABL rule. For as we said in comment 2 above, only the part of $S$ outside the light cone of a given point $y$ is considered. The outcomes of (notional) measurements in the causal future of $y$ are not conditioned on, as they would be in a straightforward relativistic generalization of the ABL rule. However, the rule shows the legitimacy of conditioning on final as well as initial information; and is similar in spirit to Kent's proposal.

\subsection{Toy model: single ``photon" interaction}

We now describe, with a simple semi-relativistic model in one spatial dimension, a photon interacting with a non-localized system, and later registering on a late hypersurface. We will see how Kent's invocation of a final boundary condition  determines a single quasiclassical history of the relevant beable: here, the location of the system. 

Consider a system initially in a superposition of two localized states:
\begin{equation}
\psi^{\rm sys}_0 =   a \psi^{\rm sys}_1 + b \psi^{\rm sys}_2
\end{equation}
where $\psi^{\rm sys}_i$ are states localized around the points $x = x_i$, with $x_2 > x_1$. We take $ | x_1 - x_2 | $ to be large, so that there is negligible overlap in the wave functions, and we neglect any dynamics within and between the two systems.

For simplicity, ``photons" are treated as pointlike particles following lightlike segments of spacetime. Suppose such a photon, initially unentangled with the system, propagates rightwards from the direction $x = - \infty$, so that in the absence of any interaction it would reach $x= x_1 $ 
at $t=t_1$ and $x= x_2$ at $t= t_2 = t_1 + (x_2 - x_1 )$. We take the ``photon''-system interaction to instantaneously reverse the ``photon'''s direction of travel, while leaving the system unaffected. Cf. Figure 3.

\begin{figure}[h]
\center
\includegraphics[width=\textwidth]{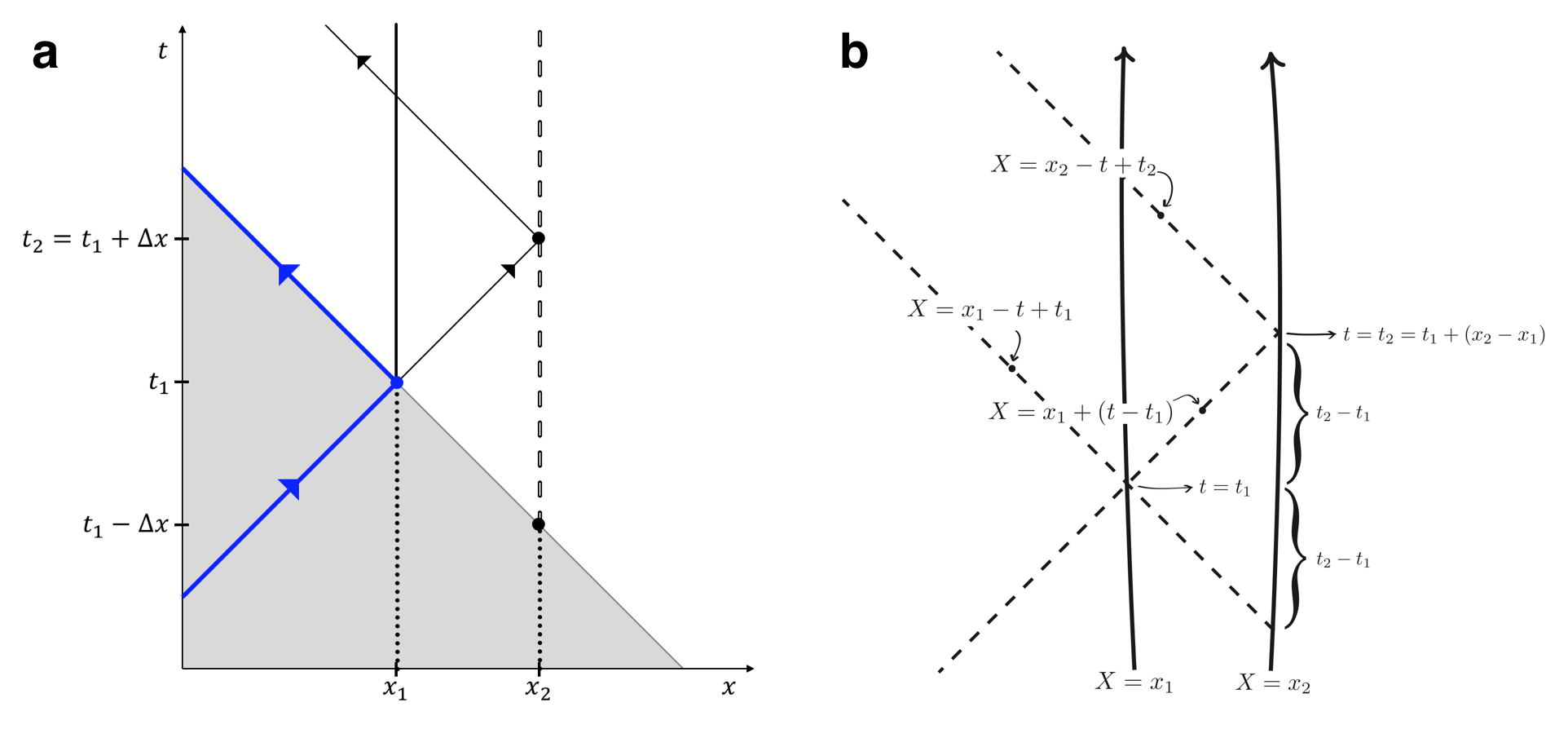}
\caption{\textbf{a}: Single ``Photon'' Interaction. Diagonal lines represent possible ``photon'' paths, with the thick blue lines representing the actual path taken. The grey region depicts the {\em region of indeterminacy}, in which the final boundary condition is uncorrelated with the beables. The vertical lines at $x_1$ and $x_2$ show the temporal structure of the collapse of the wave function, and $\Delta x = x_2-x_1$. Dotted lines represent partial mass-energy, solid vertical lines represent full mass-energy, and dashed lines represent no mass-energy. \textbf{b}: A fully labelled diagram showing the key trajectories and time intervals.}
\end{figure}

The possible outcomes of a (fictitious) mass-energy measurement at a
late time $t = T \gg t_2$ are thus either: (i) finding the photon heading
along the first leftward ray $ X = x_1 + t_1 - t$ (cf. Figure 3b) and the system accordingly localized in
the support of $\psi^{\rm sys}_1$; or (ii) finding the photon heading along
the second leftward ray $X= x_2 + t_2 - t$ and the system accordingly localized in the
support of $\psi^{\rm sys}_2$.

Suppose the real world is defined by the first outcome, (i). That is: ``Nature selects'' a final boundary condition with the photon being detected where the late-time hypersurface intersects the first leftward ray $ X = x_1 + t_1 - t$.

Then the conditional probability of finding any mass localized around the \textit{other} massive subsystem, at $x_2$, becomes zero. The beables thus describe zero mass around $x_2$, in our frame, at all the times $t$ for which the photon hits (while flying leftwards) the final hypersurface defined by the time slice $t = T$ outside the future light cone of $(x_2, t)$.

It is only after $t_1$, i.e. when $t > t_1$, that the event of hitting the final hypersurface is outside the  future light cone of $(x_1, t)$---thus the beables ``collapse" to the full mass of a single cloud at $x_1$ at time $t_1$. Since the photon escapes the light cone from $x_2$ before escaping the light cone from $x_1$, the ``collapse'' to zero mass around $x_2$ happens, in our frame, before the ``collapse'' to full-mass around $x_1$ happens. Hence Figure 3a's  triangular region of indeterminacy. 

To repeat: We suppose the real world is defined by the leftmost reflection from $x_1$, as encoded in the final boundary condition. That is: the photon reflects from $x_1$ at $t_1$, and registers on the given  time slice $t = T$ (which is much later: $T \gg t_2 > t_1$). Given this outcome, there is zero mass around $x_2$, in our frame, at all the times $t$ for which the  photon hits (while flying leftwards) the time slice $t = T$ outside the future light cone of $(x_2, t)$. That is: at all times $t$ later than $2t_1 - t_2$: a semi-infinite period. Early in this period---i.e. for $t_1 > t > 2t_1 - t_2$--- the  photon  hitting the   $t = T$ time slice is still {\em inside}  the future light cone of $(x_1, t)$. It is only after $t_1$, i.e. when  $t > t_1$, that hitting the much-later slice is outside the  future light cone of $(x_1, t)$. Thus the ``collapse'' to zero mass around $x_2$ happens, in our frame, before the  ``collapse'' to full mass around $x_1$ happens.

\subsection{Toy model: two ``photon" interaction}

Adding a second leftward ``photon'' gives a similar story to that in Section 3.1's toy model. But now there is a ``collapse'' to full mass and zero mass across a light cone, instead of across a single light ray. In our 1+1-dimensional spacetime, this yields a triangular region of indeterminacy; cf. Figure 4.

Using $x$ to label the position variable for the first photon, $y$ for the system, and $z$ for the second photon: the wave-function for times $t<t_1$ is
\begin{equation}
	\psi(x,y,z;t) = \delta(x-(x_1+t-t_1))\cdot\delta(z-(x_2-t+t_1))\cdot(a\psi_1(y)+b\psi_2(y)) \; .
\end{equation}

Again, we suppose Nature selects a late-time mass-energy distribution indicating that both photons  reflected from mass cloud one, i.e. at  $x_1$. Then, as in the first toy model:

\vspace{0.2cm}
\noindent (i) the leftmost outgoing light ray for the first photon reflecting off mass cloud one encodes all the relevant information about the system;

\vspace{0.2cm}
\noindent (ii)  the other light ray for the first photon reflecting off mass cloud two is inconsequential. It lags behind the leftmost light ray, thus manifesting itself at a later time in the conditional probability. 

\begin{figure}[h!]
\center
    \includegraphics[scale=0.35]{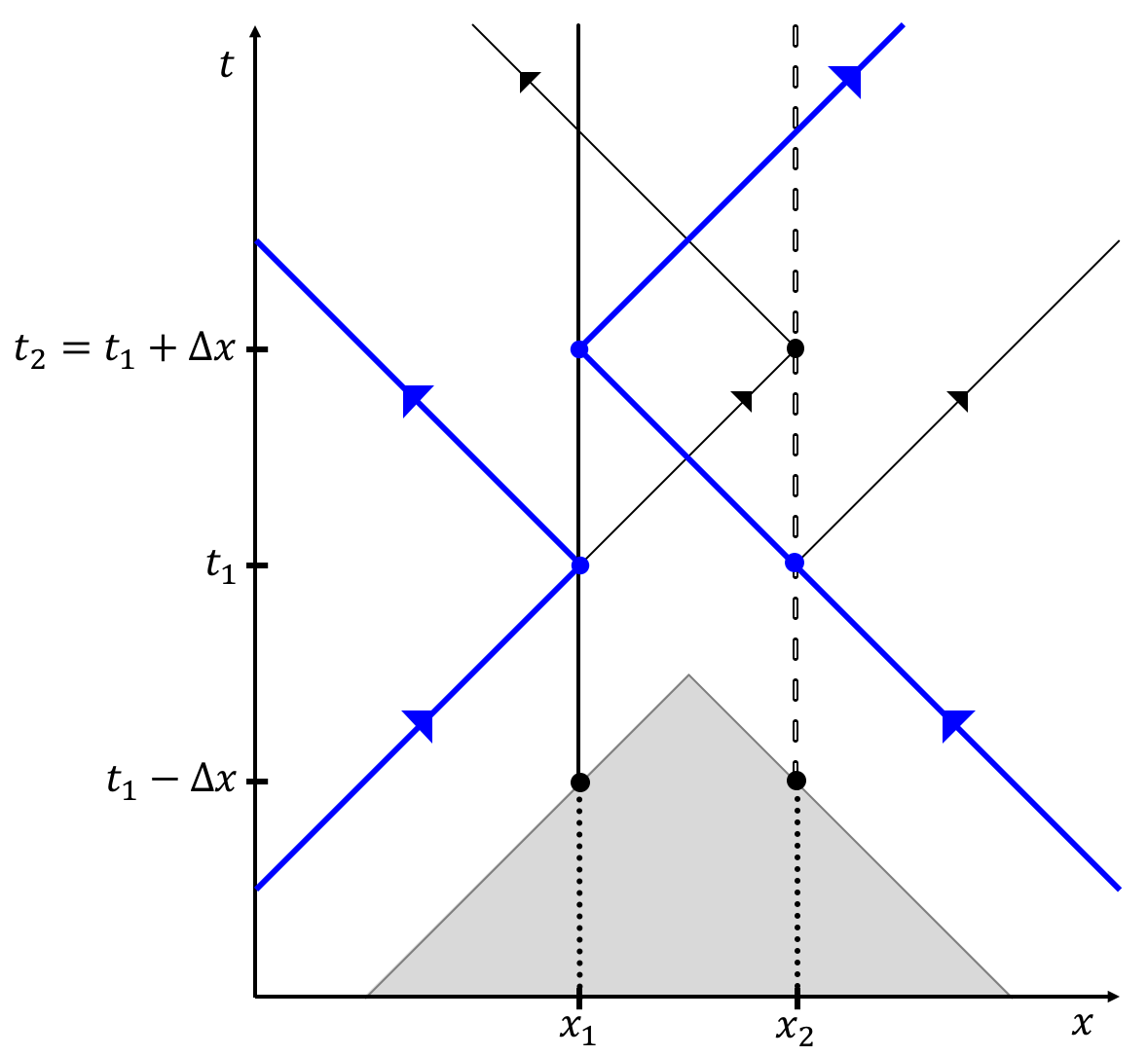}\hspace{2pc}%
\begin{minipage}[b]{14pc}\caption{\label{TwoPhotons}Two ``Photon'' Interaction. The same graphical representation, but now with two photons approaching from $\pm$ spatial infinity and a resulting triangular region of indeterminacy.}
\end{minipage}
\end{figure}

However, the rightmost outgoing ray for the second (initially-leftward) photon reflecting from mass cloud two also encodes all the relevant information about the system, in the sense that: if mass-energy is found along this light ray, the system must be found around mass cloud two---otherwise, it must be found around mass cloud 1.

Thus, we now have a ``collapse'' along the rightmost outgoing light ray as well as the leftmost outgoing light ray. So for any point  $y$ in spacetime which is to the future of at least one of these light rays, the relevant information about  localizing the  cloud is in the late-time mass-energy distribution in the surface $S_{out}(y)$ of Figure 2. So the region of spacetime which does {\em not} ``see the collapse'' must lie to the past of both  the leftmost and the rightmost outgoing light rays---resulting in a triangular region of indeterminacy.

\section{Non-locality in Kent's proposed formulation}\label{CRLL}
In this Section, we turn to the question: does Kent's proposal satisfy Shimony's conditions discussed in Section 2, viz. Outcome Independence (OI) and Parameter Independence (PI)? In short, we will find that Kent's proposal, once adapted to a Bell experiment, is much like the pilot-wave theory. (Our full analysis is in \cite{Butterfield}.) 

Regarding Outcome Independence,  Kent's proposal obeys OI at the `micro-level', and outcome dependence (OD), i.e. quantum orthodoxy, at the `observable level' (Sections 4.1 and 4.2). Here,  `micro-level' and `observable level' have meanings analogous to their meanings in pilot-wave theory. So the idea is that the `micro-level' includes all information: in pilot-wave theory, it includes the point-particles' positions (as well as the quantum state); and in Kent's proposal, it includes the final condition, i.e. the result of the notional measurement on the final hypersurface. On the other hand, in both theories---pilot-wave theory and Kent's proposal---`observable level' refers to orthodox probabilities prescribed by just the quantum state.

Regarding Parameter Independence, one needs to consider the recent important `no-go' theorems of Colbeck-Renner, Landsman, and Leegwater. The gist of these theorems is that a theory supplementing orthodox quantum theory must violate: {\em either}
\begin{enumerate}
    \item Probabilities of hidden variables are independent of which quantity is measured: this is usually labelled `no conspiracy', and so it seems mandatory: but as we said at the end of Section 2 (after equation 4) and in comment 3 of Section 3, the label is tendentious, since there need be nothing conspiratorial---or in any way suspicious---about conditioning on future information that contains traces (records) of earlier choices of measurement; {\em or}
    \item Parameter Independence.
\end{enumerate}
Pilot-wave theory violates (ii). We will see in Section 4.3 that at first sight, Kent's proposal is similar. But the question is open: it may violate (i) and obey PI. 

\subsection{Outcome Independence at the `micro-level'}
We take an outcome to be encoded by a ``photon'' registering on the surface $S$. Compare how in the pilot-wave theory, an outcome of a spin measurement by a Stern-Gerlach magnet is encoded by  the point-particle's position. 

We  adapt Kent's first toy model (Section 3.1) to describe a Bell experiment by, firstly, taking the locations $x_1$ and $x_2$ ($x_2 > x_1$) to be the left-wing of the experiment. We take the two possible outcomes of a measurement on the system entering the left wing (`the L-system') to be localization of mass density around $x_1$ and $x_2$. In any run of the experiment, one outcome is definite, thanks to reflection of an initially-rightward ``photon''. (Alternative settings are a topic only for Parameter Independence.)

Similarly, there is a right wing and an R-system, which reflects an initially-leftward ``photon''. The initial joint  state  (with $Y_L, Y_R$ for the two systems' spatial coordinates) is:  
\be 
a \psi_1(Y_L)\psi_4(Y_R) + b \psi_2(Y_L)\psi_3(Y_R)
\label{jointtoy}
\ee
where $x_3$ and $x_4$ are located far along the $x$-axis, i.e. $x_3 >> x_1, x_2$ and $x_4  >> x_1, x_2$, and with $x_3 < x_4$.
 
Suppose that in one actual run of the experiment, the `outer', $x_1$ and $x_4$, outcomes occur in the quasiclassical history specified by the final boundary condition, which we write as $t_S$. That is: (a) a photon registers on $S$ so far to the left as to imply that it earlier reflected at $x_1$ rather than at $x_2$, while (b) another photon registers on $S$ so far to the right as to imply that it earlier reflected at $x_4$ rather than at $x_3$.

If we condition orthodox quantum probabilities for outcomes on all this (very rich!) information  in $t_S$, the probabilities will become trivial, i.e. 0 or 1.  And so they will factorize. So we obtain probability 1 for each of the two `outer', $x_1$ and $x_4$, outcomes. The situation is similar to the way that, once particle positions are take into account, the pilot-wave theory is past-deterministic just as much as it is future-deterministic; and so one could discuss its satisfaction of OI by invoking  particles' {\em final} positions, instead of (as usual) their initial ones.

\subsection{Outcome dependence at the observable level}
Of course, we do not know the final condition $t_S$;  and indeed, we will never know more than a tiny fraction of the actual single quasiclassical history. Kent's proposal about this is natural (and again analogous to pilot-wave theory):

\begin{enumerate}
    \item One should average over the possible final mass-energy distributions $t_S$ on  $S$,  with their Born-rule probabilities as prescribed by the final quantum state $| \psi_S \rangle$.
    \item It is these averaged probabilities that are equal, or close enough, to the orthodox probabilities.
\end{enumerate}

The analogy with the pilot-wave theory's averaging over its (necessarily unknown!) particle positions, using their Born-rule probabilities, is clear. So as for the pilot-wave theory: in so far as we are confident that Kent's proposal recovers orthodox quantum probabilities at the observable level, we thus conclude that it  satisfies OD at that level.

\subsection{Parameter Independence?}
Assessing PI, at either the micro or the observable levels, is a substantive task. One needs to model in a Kentian fashion both:

\begin{enumerate}
    \item different choices of apparatus-setting (parameter) on one wing, and
    \item non-selective measurement, (i.e. defining the left-wing marginal probability by summing probabilities of the various right-wing outcomes).
\end{enumerate}

A Kentian model of (i) will involve photons  first scattering off the apparatus' knob that sets which quantity gets measured, and then much later, registering on the hypersurface $S$; so that the actual final condition encodes the setting. As we emphasized in (i) at the start of this Section: a final condition encoding a knob-setting, and probabilities dependent on such a setting, is not {\em ipso facto} suspicious. To model (ii), the probabilities dependent on reflection from ``which knob-setting'' will need to be combined with probabilities  dependent on reflection from ``which outcome, i.e. mass-energy lump''.

We leave the development of a model addressing both (i) and (ii), to another paper. But whatever its details, one will need to consider how recent `no go' theorems about non-locality bear on it. We will now briefly discuss this. 

Colbeck and Renner \cite{CR} (made rigorous by Landsman \cite{Landsman}) show that, under some apparently natural extra assumptions: any theory that supplements orthodox quantum theory, in the sense of recovering orthodox Born-rule probabilities by probabilistic averaging over ``hidden variables'',  must violate either:
\begin{enumerate}
    \item Probabilities of hidden variables are independent of which quantity is measured: called `no conspiracy'---tendentiously---cf. equation (\ref{recoverbyaverage}), or
    \item Parameter Independence.
\end{enumerate}
Leegwater \cite{Leegwater} dispenses with the extra assumptions apart from a `no conspiracy' assumption: that the measure used to average over the hidden variables so as to recover Born-rule probabilities must be independent of which quantity, or quantities, are (chosen to be) measured. 

Since Kent's probabilities are conditioned on a specific final boundary condition, they are not equal to orthodox probabilities, and thus his proposal must violate (i) or (ii).  
At first sight, it seems that Kent's proposal satisfies no conspiracy. For he averages over the possible final boundary conditions, with the Born-rule probabilities given by the final universal quantum state  $| \psi_S \rangle$. And $| \psi_S \rangle$ is independent of which quantity, or quantities, were previously measured, even though it is the universal state.

 Agreed: there is a fact about which quantity got measured (and about what the outcome was), according to Kent's overall scheme for recovering a quasiclassical world. But these facts leave no `mark' on---have no back-reaction on---the universal quantum state, which always evolves unitarily. (This is like the pilot-wave theory's particle being guided by the $\psi$-wave, but having no influence on how $\psi$ evolves.) Thus in the Schr\"{o}dinger picture, $| \psi_S \rangle$ gets, at an intermediate time when an experiment is set up, components (non-zero amplitudes) for various possible choices of quantity (settings) and, soon thereafter, components  for various possible outcomes. (Think of an Everettian's description of the setting-up, and performance, of a run of an experiment: the components correspond to the Everettian's branches.)
 
 But on reflection, it is not clear that Kent's proposal satisfies no conspiracy. For the no conspiracy assumption concerns the measure used in deriving a Bell inequality. But {\em a priori}, this measure need not be the Born-rule probabilities given by the final universal quantum state; (cf. Section 5.1 of \cite{Butterfield}). This means that on the question whether Kent's proposal satisfies Parameter Independence, the jury is still out.

\section{Towards alternative proposals}\label{Marsh}

Kent's proposal requires that photons interact with systems in superposed states, and so become entangled with the systems, and appropriately encode one or another component of the superposition---and then escape to spatial infinity in order for a quasiclassical history to emerge. So, for example, the beables for a quantum system in a superposed state which is completely contained at the centre of a large photon-absorbing box could never describe the emergence of a quasiclassical history. 

Moreover, perhaps many quantum systems are not so ``fortunate'' as to interact with photons that then escape unimpeded to infinity. For example, photons that through entanglement record the state of macroscopic systems  might only escape to infinity after several scattering events. For such systems, Kent's approach will imply an {\em information delay}: a time-lag until ``collapse'', which occurs after the point of interaction.

This situation prompts us to explore modifying the Lorentz-invariant rule for constructing beables, so as to avoid any information delay. The rest of this Section builds from our ideas in \cite{Marsh2018,Marsh2019} toward this goal.



\begin{figure}[h]
\center
\includegraphics[scale=0.9]{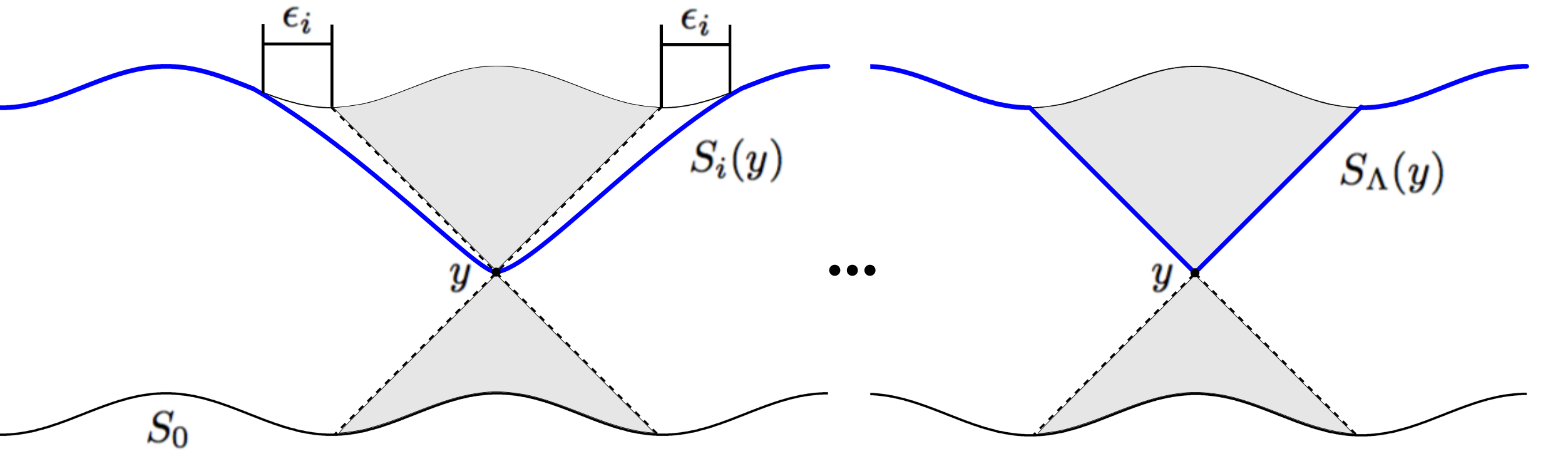}
\caption{Spacelike hypersurfaces $S_i(y)$ (left) converging, as $\epsilon_i\to 0$, to the {\em effective future boundary} $S_\Lambda(y) $ (right) for a given spacetime point $y$ between $S_0$ and $S$.}
\label{Si}
\end{figure}

We begin by noting that in Kent's original proposal, beables at the spacetime point $y$ are constructed via conditional expectation values for simultaneous measurements of all points $x$ in some set $S_i \cap S$, where $\{S_i\}_i$ defines a sequence of hypersurfaces converging to the hypersurface $S_\Lambda(y)$, as defined in figure \ref{Si}. The probabilities are conditional on a final boundary condition giving the actual mass-energy on the final hypersurface $S$.

We now suggest: i) defining a sample space of mass-energy distributions on the entire spacetime, not just the final hypersurface; ii) using these distributions to formulate conditional probabilities for a point $y$ which are conditional upon (notionally!) measuring the actual mass-energy distribution at \textit{all} points $x\in S_i$, not just $x\in S_i \cap S$. So in the limit as $i\to\infty$, this effectively allows the conditional probabilities to be sensitive to mass-energy on the forward light cone of $y$, not just outside of it. Again, cf. figure \ref{Si}. 

\begin{figure}[h]
\centering
\hspace*{-1cm}
    \includegraphics[width=15cm]{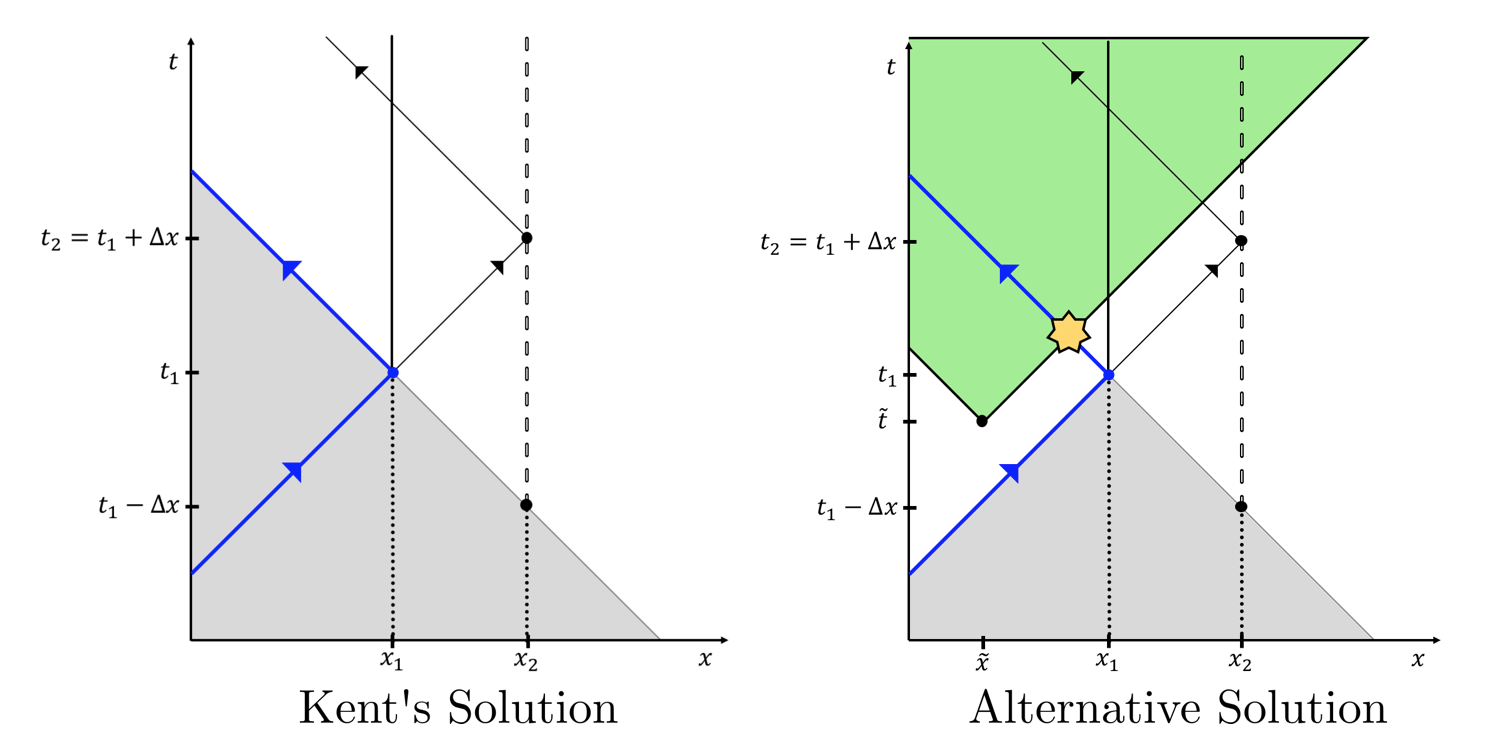}
    \caption{The  single ``photon'' toy-model in the new approach compared to Kent's proposal. For the point $(\tilde x,\tilde t)$, its forward light cone, with green interior, contains pings of mass energy from outgoing ``photons''.  So the actual mass-energy distribution \textit{could} contribute to determining beables at points such as $(\tilde x,\tilde t)$, whereas in Kent's approach it would not. For this particular system, while the region of indeterminacy shrinks, the collapse of beables is described  in exactly the same way. }
\label{AltSoln}

\end{figure}

Intuitively, this allows the beables to be sensitive to entangled photons crossing the boundary of their light cone, registering ``pings'' of energy there, in addition to entangled photons outside the light cone. This subtle addition would have significant ramifications for the beables describing physical reality, and in particular, information delay would be avoided.

In Section 3.1's single ``photon'' toy model, we can easily identify two possible mass-energy histories for the whole spacetime, as opposed to just the final boundary conditions. These are, similarly to Section 3.1: (i) the mass-energy distribution of the photon in the scenario where it reflects from $x_1$, along with the mass-energy of the massive subsystem localized at $x_1$; and (ii) the mass-energy distribution of the photon in the scenario where it reflects from $x_2$, along with the mass-energy of the massive subsystem localized at $x_2$.

The only spacetime points $y$ for which the light cone of $y$ encodes the mass-energy distribution while the hypersurface $S_{out}(y)$ does not---thus implying a difference from Kent's proposal---is the region to the left of $x_1$ with time coordinate greater than the incoming photon but less than the outgoing photon. At these points, the outgoing photon produces a ping on the forward light cone, thus revealing the presence or absence of the photon there in the mass-energy distribution. So the region of indeterminacy shrinks as compared to Kent's proposal: cf. figure \ref{AltSoln}. But the beables themselves do not change in this model. In this sense, the predictions of our suggested approach coincide, for this system in which no information delay is present, with Kent's predictions.

\begin{figure}[h]
\centering
\hspace*{-1cm}
\includegraphics[scale=0.9]{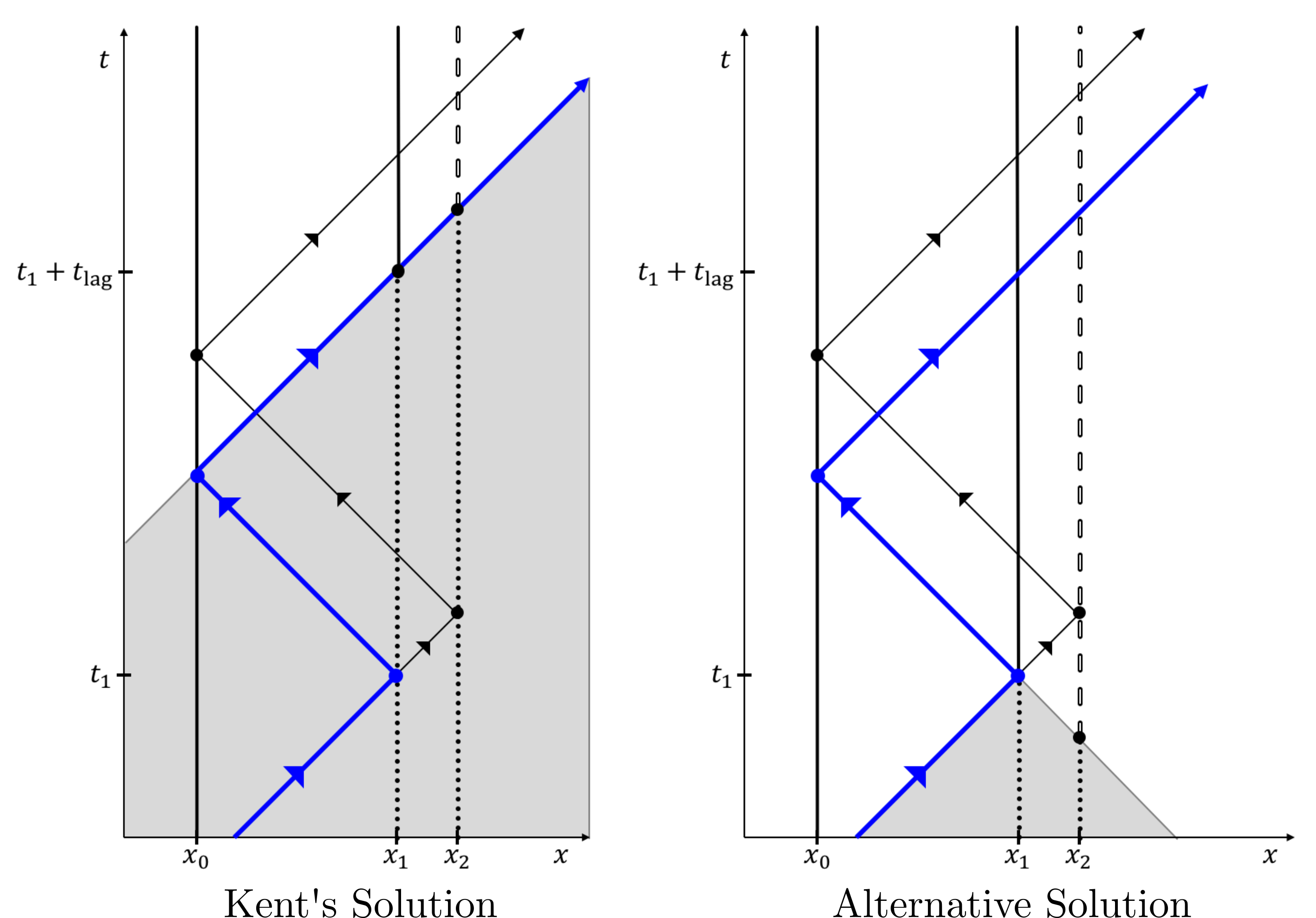}
\caption{A comparison of Kent's approach and the alternative, for a scenario with information delay. In the alternative approach, the ``superposed" state of the beables collapses, before or precisely at the time of interaction $t_1$.  }

\label{NoLag}
\end{figure}

We can also compare the approaches in a scenario for which, on Kent's approach, information delay \textit{is} present. This can be done with a simple adaptation of the single ``photon'' toy model. Suppose there is another massive body, fixed at $x_0\ll x_1$, so that we can ignore any interaction between the massive body and the superposed system. We introduce the photon from a position to the left of $x_1$ but to the right of $x_0$, and suppose that the photon will reach $x_1$ at some time $t_1$. Cf. figure \ref{NoLag}.

After reflecting from the first or second mass cloud, the photon reverses direction toward $x_0$ and reflects again upon reaching $x_0$. Technically, the photon should be trapped in the region between $x_0$ and $x_1$ or $x_2$ forever after reflecting from the system in superposition. However, this is clearly just an artifact of working in one spatial dimension; and we can avoid this artifact, by instead supposing the photon passes unimpeded through $x_1$ and $x_2$ upon returning for the second time. The important difference from the original toy model is that the photon undergoes one further interaction after reflecting from the system in superposition, before propagating toward spatial infinity to the right.

We again suppose that the late-time beable configuration chosen by Nature indicates that the photon reflects from the first mass cloud, at position $x_1$. Applying Kent's rules, there is indeed information delay: cf. figure \ref{NoLag} (left-hand side). The outgoing photon does not appear outside the light cone of $x_1$ at the time of interaction $t_1$, but at a later time $t_1+2|x_1-x_0|/c$.

In this scenario, conditioning on the mass-energy on the future light cone provides far more information, since pings of mass-energy are registered on the light cone, as the ``photon'' bounces around. The region of indeterminacy also shrinks. The superposition collapses at time $t_1$ around $x_1$, and at time $t_1-|x_2-x_1|/c$ around $x_2$ as shown in figure \ref{NoLag} (right-hand side).

An immediate challenge for making our suggested approach rigorous is defining a probability measure of mass-energy distributions in spacetime. For we can no longer simply appeal to the Born rule on the final hypersurface. Assuming this technical challenge could be met, such a modification to Kent's framework could eliminate the information delay phenomenon.

\section{Conclusion}\label{summary}

In this paper, we have analyzed Kent's proposal for a Lorentz-invariant solution to the measurement problem from a variety of perspectives. We summarize with four points:

\begin{enumerate}
    \item Shimony hoped for a peaceful coexistence between quantum non-locality and special relativity by ``blaming'' the Bell-inequality derivation on assuming Outcome Independence, {\em pace} the pilot-wave theory: a viewpoint that, we urged, makes the measurement problem vivid.
    \item Kent has proposed a solution to the problem that is: realist, one-world, relativistic, and without state reduction. The idea is: a final boundary condition makes mass-energy definite at intermediate times, via scattered photons, and this yields a single actual quasiclassical history.
    \item Kent's proposal is like the pilot-wave theory in that: it obeys OI at the micro-level (i.e. with the final boundary condition specified)---but presumably, not at the observable level. Theorems by Colbeck, Landsman, Leegwater, and Renner imply that it also violates either: ``no conspiracy'', or Parameter Independence at the micro-level.
    \item When none of the photons scattered from a macro-event (that we intuitively want to be definite) escape uninterrupted to spatial infinity from the macro-event, a quasiclassical history is, on Kent's proposal, \textit{not} recovered. We outlined a revision. The idea is to condition on mass-energy present on the future light cone, as well as on mass-energy in the final boundary condition outside the future light cone.
\end{enumerate}

\ack{We thank Adrian Kent for comments and conversations, especially about Section 5's suggested extensions of his proposal; and Gijs Leegwater for informative discussions about quantum non-locality. JB also thanks Bryan Roberts for discussions and Figure 3b. BM also acknowledges support from the National Science Foundation Graduate Research Fellowship under Grant No. DGE - 1656518.}

\section*{References}
\bibliography{biblio}

\end{document}